\documentclass[preprint,amsmath,amssymb,aps,showkeys,showpacs]{revtex4}
\usepackage[english]{babel}
\usepackage{graphicx}
\usepackage{graphics}
\usepackage{amsmath}
\usepackage{dcolumn}
\usepackage{amssymb}
\usepackage{bm}

\begin{document}
\title{Non-relativistic quantum effects on the harmonic oscillator in a spacetime with a distortion of a vertical line into a vertical spiral }
\author{W. C. F. da Silva}
\affiliation{Departamento de F\'isica, Universidade Federal da Para\'iba, Caixa Postal 5008, 58051-900, Jo\~ao Pessoa, PB, Brazil.}

\author{R. L. L. Vit\'oria}
\email{ricardo.vitoria@ufes.br}
\affiliation{Departamento de F\'isica e Qu\'imica, Universidade Federal do Esp\'irito Santo, Av. Fernando Ferrari, 514, Goiabeiras, 29060-900, Vit\'oria, ES, Brazil.}

\author{K. Bakke}
\email{kbakke@fisica.ufpb.br}
\affiliation{Departamento de F\'isica, Universidade Federal da Para\'iba, Caixa Postal 5008, 58051-900, Jo\~ao Pessoa, PB, Brazil.}

\begin{abstract}

Non-relativistic quantum effects of the topology of the spacetime with the distortion of a vertical line into a vertical spiral on the harmonic oscillator are investigated. By searching for analytical solutions to the Schr\"odinger equation in this topological defect background, it is shown that the topology of the spacetime modifies the spectrum of energy of the harmonic oscillator. Besides, it is shown that there exists an Aharonov-Bohm-type effect for bound states.

\end{abstract}


\maketitle

\section{Introduction}

Topological defect spacetimes have brought a great attention in recent decades. The best example of a topological defect spacetime is the cosmic string spacetime \cite{AV,vil,JSD,kibble}. It has a singularity determined by the curvature concentrated on its symmetry axis. It is known as a conical singularity \cite{staro}. Interesting works \cite{mm,corda2,corda3,corda7,corda9,corda12,corda13,godel1,godel2,godel3,do,kgo} have explored this topological characteristic of the cosmic string with the purpose of searching for analogues of the Aharonov-Bohm effect \cite{valdir,valdir2}. It is worth clarifying that Peshkin and Tonomura \cite{pesk} showed that if a quantum particle is confined to move in a circular ring of radius $R$ and there is a solenoid (extremely long) of radius $a<R$, then, the angular momentum quantum number is modified by $l_{\mathrm{eff}}=l-e\Phi/2\pi$ (where $\Phi$ is the magnetic flux through the solenoid and $e$ is the electric charge). In this way, the spectrum of energy becomes determined by $l_{\mathrm{eff}}$ even though no interaction between the particle and the magnetic field inside the solenoid exists. This influence of the magnetic flux on the energy levels is known as the Aharonov-Bohm effect for bound states. Hence, an analogue effect of the Aharonov-Bohm effect that we have mentioned corresponds to the influence of the topological defect on the spectrum of energy even though no interaction between the quantum particle and the topological defect exists. Another interest is the topological defect spacetimes associated with torsion, since Aharonov-Bohm-type effects have been reported in the literature. Examples of Aharonov-Bohm-type effects are given by relativistic quantum particles in the spacetime with a space-like dislocation \cite{valdir4,bf3,bf4,vb2,vb3}, the chiral conical spacetime \cite{valdir3,gal,fur4} and the spacetime with a spiral dislocation \cite{bf2,vb}.

Another perspective of searching for Aharonov-Bohm-type effects due to backgrounds of topological defect spacetimes was given in the non-relativistic limit. By dealing with the non-relativistic limit of the Klein-Gordon equation and the Dirac equation in topological defect spacetimes, therefore, it has been shown the connection with the elastic theory in solids \cite{kleinert,kat}. Since then, the spatial part of the line element of a topological defect spacetime describes defects in solids, such as, disclinations and dislocations. In particular, examples of dislocations are the screw dislocation and the spiral dislocation \cite{val,put}. Quantum effects associated with a screw dislocation have been studied in quantum rings \cite{fur6,fur3,dantas1}, electrons subject to the deformed Kratzer potential \cite{1}, electron gas in a cylindrical shell \cite{shell}, and with an electron in a uniform magnetic field \cite{fur,fur2,fur7,fil}. On the other hand, quantum effects associated with a spiral dislocation have been studied with geometric quantum phases \cite{bf2} and the harmonic oscillator \cite{mb}. Other interesting works that follow this line of research are \cite{fur8,corda6,corda8,corda10,corda11}.

In this work, we follow this line of study that searches for Aharonov-Bohm-type effects in the non-relativistic limit. We analyse non-relativistic quantum effects on the harmonic oscillator in the spacetime with the distortion of a vertical line into a vertical spiral. Then, we show that analytical solutions to the Schr\"odinger equation for the harmonic oscillator in this background can be obtained. Due to the effects of the topology of the spacetime, the spectrum of energy of the harmonic oscillator is modified. Besides, we discuss the appearance of an Aharonov-Bohm-type effect for bound states.

This paper is structured as follows: in section II, we introduce the line element of the spacetime with the distortion of a vertical line into a vertical spiral. Then, we analyse the topological effects on the harmonic oscillator by show that an analogue of the Aharonov-Bohm effect for bound states exists; in section III, we present our conclusions.

\section{Spacetime with a distortion of a vertical line into a vertical spiral}

Let us start by considering a generalization of a topological defect in gravitation. From Ref. \cite{val}, we can write the line element of a spacetime with a distortion of a vertical line into a vertical spiral as
\begin{eqnarray}
ds^{2}=-c^{2}dt^{2}+dr^{2}+r^{2}d\varphi^{2}+2\beta\,d\varphi\,dz+dz^{2},
\label{1.1}
\end{eqnarray}
where $0\,<\,r\,<\,\infty$, $0\leq\varphi\leq2\pi$ and $-\infty\,<\,z\,<\,\infty$. The parameter $\beta$ is a constant that characterizes the torsion field (dislocation). As mentioned in the introduction, there is a connection with the elastic theory in solids \cite{kleinert,kat} when we consider the spatial part of the line element of a topological defect spacetime to describe a defect in solids. As shown in Ref. \cite{val}, the parameter $\beta$ is related to the Burger vector $\vec{b}$ ($\beta\propto\left|\vec{b}\right|$). Thus, the spatial part of the line element (\ref{1.1}) describes a kind of screw dislocation, since it has the Burger vector perpendicular to the plane $z=0$. Besides, the value of the parameter $\beta$ in a solid is of the order of the interatomic displacement, hence, it can be defined in the range $0\,<\,\beta\,<\,1$. Therefore, in terms of the elastic theory in solids, this screw dislocation corresponds to the distortion of a vertical line into a vertical spiral \cite{val}. Note that it differs from the screw dislocation worked in Refs. \cite{fur6,fur3,dantas1,fur8,fur}.

Since our aim is to search for Aharonov-Bohm-type effects in the non-relativistic limit. In particular, we are going to analyse the topological effects of the the distortion of a vertical line into a vertical spiral on the two-dimensional harmonic oscillator. For this purpose, let us follow Refs. \cite{fur,fur2,fur7}, then, the time-independent Schr\"odinger equation for the harmonic oscillator in the presence of a topological defect is written in the form: 
\begin{eqnarray}
\mathcal{E}\psi=-\frac{\hbar^{2}}{2m}\frac{1}{\sqrt{g}}\,\partial_{k}\left[\sqrt{g}\,g^{kj}\,\partial_{j}\right]\psi+\frac{1}{2}m\,\omega^{2}r^{2}\psi,
\label{1.2}
\end{eqnarray}
where $g_{ij}$ is the metric tensor, $g^{ij}$ is the inverse of $g_{ij}$ and $g=\mathrm{det}\left|g_{ij}\right|$. Observe that the indices $\left\{i,\,j\right\}$ run over the space coordinates. Henceforth, we work with the units: $c=1$ and $\hbar=1$.

Thereby, with the line element (\ref{1.1}), the time-independent Schr\"odinger equation (\ref{1.2}) becomes
\begin{eqnarray}
\mathcal{E}\psi&=&-\frac{1}{2m}\left[\frac{\partial^{2}\psi}{\partial r^{2}}+\frac{r}{\left(r^{2}-\beta^{2}\right)}\,\frac{\partial\psi}{\partial r}+\frac{1}{\left(r^{2}-\beta^{2}\right)}\left[\frac{\partial}{\partial\varphi}-\beta\frac{\partial}{\partial z}\right]^{2}\psi+\frac{\partial^{2}\psi}{\partial z^{2}}\right]+\frac{1}{2}m\omega^{2}r^{2}\psi.
\label{1.3a}
\end{eqnarray}

Due to the cylindrical symmetry determined by the spacetime line element (\ref{1.1}), it is possible to write the solution to Eq. (\ref{1.3a}) as $\psi\left(r,\,\varphi,\,z\right)=e^{il\varphi+ikz}\,R\left(r\right)$, where $k=\mathrm{const}$ and $l=0,\pm1,\pm2,\pm3\ldots$ are the eigenvalues of the operators $\hat{p}_{z}=-i\partial_{z}$ and $\hat{L}_{z}=-i\partial_{\varphi}$, respectively. Thereby, from  Eq. (\ref{1.3a}), we obtain the radial equation:
\begin{eqnarray}
R''+\frac{r}{\left(r^{2}-\beta^{2}\right)}\,R'-\frac{\left(l-\beta\,k\right)^{2}}{\left(r^{2}-\beta^{2}\right)}\,R-m^{2}\omega^{2}r^{2}\,R+\left(2m\mathcal{E}-k^{2}\right)\,R=0.
\label{1.3}
\end{eqnarray}

From now on, let us define the parameters:
Let us call
\begin{eqnarray}
\gamma=l-\beta\,k;\,\,\,,\tau=2m\mathcal{E}-k^{2},
\label{1.4}
\end{eqnarray}
and thus, by performing the change of variables $x=r^{2}/\beta^{2}$, the radial equation (\ref{1.3}) becomes
\begin{eqnarray}
4xR''+\frac{\left(4x-2\right)}{\left(x-1\right)}\,R'-\frac{\gamma^{2}}{\left(x-1\right)}R-m^{2}\omega^{2}\beta^{4}x\,R+\beta^{2}\tau\,R=0.
\label{1.5}
\end{eqnarray}

Next, let us analyse the behaviour of the radial wave function when $x\rightarrow\infty$. In this way, the solution to Eq. (\ref{1.5}) is given by
\begin{eqnarray}
R\left(x\right)=e^{-\frac{m\omega\beta^{2}}{2}\,x}\,f\left(x\right),
\label{1.6}
\end{eqnarray}
where $f\left(x\right)$ is an unknown function. Then, by substituting Eq. (\ref{1.6}) into Eq. (\ref{1.5}), we obtain the following differential equation for $f\left(x\right)$: 
\begin{eqnarray}
f''&+&\left[\frac{1}{\left(x-1\right)}-\frac{1}{2x\left(x-1\right)}-m\omega\beta^{2}\right]f'-\frac{\gamma^{2}}{4x\left(x-1\right)}f-\frac{m\omega\beta^{2}}{2\left(x-1\right)}f+\frac{m\omega\beta^{2}}{4x\left(x-1\right)}f\nonumber\\
&+&\frac{\beta^{2}\tau}{4x}f=0.
\label{1.7}
\end{eqnarray}

Hence, Eq. (\ref{1.7}) is known in the literature as the confluent Heun equation \cite{heun}, and thus the function $f\left(x\right)$ is the confluent Heun function: 
\begin{eqnarray}
f\left(x\right)=H_{\mathrm{C}}\left(-m\omega\beta^{2},\,-\frac{1}{2},\,-\frac{1}{2},\,\frac{\beta^{2}\tau}{4},\,\frac{3}{8}-\frac{\gamma^{2}}{4}-\frac{\beta^{2}\tau}{4};\,x\right).
\label{1.8}
\end{eqnarray}

With the aim of dealing with the confluent Heun equation (\ref{1.7}), let us use the Frobenius method \cite{arf,griff}. Thereby, we can write $f\left(x\right)=\sum_{k=0}^{\infty}a_{k}\,x^{k}$, and then, from Eq. (\ref{1.7}) we obtain the relation:
\begin{eqnarray}
a_{1}=-\frac{\left[\beta^{2}\left(\tau-m\omega\right)+\gamma^{2}\right]}{2}\,a_{0},
\label{1.9}
\end{eqnarray}
and also the recurrence relation:
\begin{eqnarray}
a_{k+2}=\frac{2\left[\left(k+1\right)\left(k+1+m\omega\beta^{2}\right)-\frac{\beta^{2}}{4}\left(\tau-m\omega\right)-\frac{\gamma^{2}}{4}\right]}{\left(k+2\right)\left(2k+3\right)}\,a_{k+1}+\frac{m\omega\beta^{2}\left(\theta-4k\right)}{2\left(k+2\right)\left(2k+3\right)}\,a_{k},
\label{1.10}
\end{eqnarray}
where we have defined the parameter: $\theta=\frac{\tau}{m\omega}-2$.

A polynomial solution to the function $f\left(x\right)$ is achieved when we impose the requirement that the series terminates. The series terminates by imposing that
\begin{eqnarray}
a_{n+1}=0;\,\,\,\,\theta=4n,
\label{1.11}
\end{eqnarray}
where $n=1,2,3,\ldots$ corresponds to the quantum number related to the radial modes. Hence, the two conditions established in Eq. (\ref{1.11}) must be satisfied in order that the series terminates.

Since we are searching for bound state solutions, let us build a polynomial of first degree for $f\left(x\right)$. Thereby, let us take $n=1$. Then, from Eq. (\ref{1.11}), we have that $a_{n+1}=a_{2}=0$. By using Eq. (\ref{1.10}) and neglecting the terms of order $\mathcal{O}\left(\beta^{3}\right)$, we obtain the relation:
\begin{eqnarray}
\omega_{1,\,l,\,k}=\frac{\tau_{1,\,l,\,k}\left(\gamma^{2}-1\right)}{3m\gamma^{2}}+\frac{\left(\gamma^{4}-4\gamma^{2}\right)}{6m\beta^{2}\,\gamma^{2}}.
\label{1.12}
\end{eqnarray}

The relation (\ref{1.12}) means that the angular frequency of the harmonic oscillator has a discrete set of values that allows to achieve a polynomial of first degree to $f\left(x\right)$. For this reason we have labelled $\omega=\omega_{n,\,l,\,k}$. Therefore, the permitted values of the angular frequency associated with the radial mode $n=1$ are determined by Eq. (\ref{1.12}). For other values of $\omega_{n,\,l,\,k}$, we cannot obtain a polynomial of first degree to $f\left(x\right)$. Besides, from the condition $\theta=4n$ given in Eq. (\ref{1.11}), for $n=1$ we have
\begin{eqnarray}
\tau_{1,\,l,\,k}=6m\,\omega_{1,\,l,\,k},
\label{1.13}
\end{eqnarray}
and thus, by using Eqs. (\ref{1.12}) and (\ref{1.4}), we obtain
\begin{eqnarray}
\mathcal{E}_{1,\,l,\,k}=\frac{\left(\gamma^{4}-4\gamma^{2}\right)}{2m\,\beta^{2}\left(2-\gamma^{2}\right)}+\frac{k^{2}}{2m}.
\label{1.14}
\end{eqnarray}

Hence, Eq. (\ref{1.14}) corresponds to the permitted energies associated with the radial mode $n=1$ for a non-relativistic quantum particle confined to the harmonic oscillator in the spacetime with the distortion of a vertical line into a vertical spiral (\ref{1.1}). In contrast to the well-known spectrum of energy of the harmonic oscillator, we have that the topology of the defect modifies the spectrum of energy of the harmonic oscillator. Note that the energy levels (\ref{1.14}) depends on the parameter $\beta$, where two contributions to the allowed energies (\ref{1.14}) exist: one is the presence of the term $\beta^{-2}$, while the second contribution is the presence of the effective angular momentum $\gamma=l-\beta\,k$. In particular, this effective angular momentum is determined by a shift in the angular momentum quantum number that stems from the effects of the topology of the defect, even though no interaction of the quantum particle with the topological defect exists. Therefore, this shift in the angular momentum gives rise to an analogue of the Aharonov-Bohm effect for bound states \cite{fur8,fur6,mb,pesk}.

Furthermore, with the conditions (\ref{1.11}) satisfied, the polynomial of first degree to $f\left(x\right)$ is given by $f\left(x\right)=1+a_{1}\,x$, where we have taken $a_{0}=1$ and $a_{1}$ is given by (\ref{1.9}). In this way, the radial wave function (\ref{1.6}) becomes
\begin{eqnarray}
R_{1,\,l,\,k}=e^{-\frac{m\omega\beta^{2}}{2}\,x}\left(1+a_{1}\,x\right).
\label{1.15}
\end{eqnarray}

Finally, from Eq. (\ref{1.12}) we can see that the permitted values of the angular frequency of the harmonic oscillator are determined by the quantum numbers $\left\{n,\,l,\,k\right\}$ of the system and the parameter $\beta$ that characterizes the defect. Therefore, for other radial modes $n=2,3,4,\ldots$, we can obtain different expressions for the angular frequency $\omega_{n,\,l,\,k}$ associated with each one. As a consequence, we shall obtain new expressions for the allowed energies associated with the radial modes $n=2,3,4,\ldots$. Observe that the allowed values for $\omega_{n,\,l,\,k}$ and for $\mathcal{E}_{n,\,l,\,k}$ are obtained by following the steps from Eq. (\ref{1.9}) to Eq. (\ref{1.14}), which means that we search for a polynomial solution to $H_{\mathrm{C}}\left(x\right)$. Note that if we take the limit $\beta\rightarrow 0$ in Eq. (\ref{1.1}), hence, we recover the line element of the Minkowski spacetime. Therefore, we would have in Eq. (\ref{1.3a}) is the Schr\"odinger equation for the harmonic oscillator in the absence of defect. In this case, the solution to the radial part of the Schr\"odinger equation (\ref{1.3a}) would be given in terms of the confluent hypergeometric function. This analysis has been made in Ref. \cite{fur8}.

\section{Conclusions}

We have analysed non-relativistic effects of the topological of a spacetime with the distortion of a vertical line into a vertical spiral on the harmonic oscillator. We have shown that the Schr\"odinger equation for a quantum particle in this background can be solved analytically and, as an example, we have obtained the expression for the allowed energies associated with the radial mode $n=1$. Then, we have seen that the topology of the spacetime modifies the spectrum of energy of the harmonic oscillator. In addition, there are two contributions to the permitted energies related to the radial mode $n=1$ that arise from the topology of the spacetime. In particular, one of these contributions shows a shift in the angular momentum quantum number that gives rise to an effective angular momentum $\gamma=l-\beta\,k$. The meaning of this shift in the angular momentum is that the topology of the spacetime with the distortion of a vertical line into a vertical spiral yields an analogue of the Aharonov-Bohm effect for bound states \cite{fur8,fur6,mb,pesk}. Another point raised in our discussion is that, by searching for polynomial solutions to the confluent Heun equation, the topology of the spacetime imposes a discrete set of values for the angular frequency of the harmonic oscillator. The permitted values of the angular frequency of the harmonic oscillator are determined by the quantum numbers $\left\{n,\,l,\,k\right\}$ of the system and the parameter $\beta$ that characterizes the defect. As an example, we have obtained the expression for the permitted values of the angular frequency associated with the radial mode $n=1$. Finally, we have written the radial wave function associated with the radial mode $n=1$.

It is worth mentioning that the geometrical approach used to describe the spacetime with the distortion of a vertical line into a vertical spiral can be useful in studies of condensed matter systems. For instance, it is interesting to study the Landau quantization \cite{landau,fur,fur2}, quantum rings \cite{fur6,fur3,dantas1}, topological insulators \cite{insulator,insulator2} and neutral particle systems \cite{er,ll5,ll6} in an elastic medium with the distortion of a vertical line into a vertical spiral (\ref{1.1}).

\appendix
 
\section{Confluent Heun Equation}

The standard form of the confluent Heun equation is \cite{heunc,heunc2,heunc3,heunc4}
\begin{eqnarray}
H''+\left[\alpha+\frac{\beta+1}{x}+\frac{\gamma+1}{x-1}\right]H'+\left[\frac{\mu}{x}+\frac{\nu}{x-1}\right]H=0,
\label{a.1}
\end{eqnarray}
where $H\left(x\right)=H_{\mathrm{C}}\left(\alpha,\,\beta,\,\gamma,\,\delta,\,\eta;\,x\right)$ is the confluent Heun function. The parameters $\mu$ and $\nu$ given in the last term of Eq. (\ref{a.1}) are defined as
\begin{eqnarray}
\mu&=&\frac{\alpha}{2}\left(1+\beta\right)-\frac{\beta}{2}\left(1+\gamma\right)-\frac{\gamma}{2}-\eta;\nonumber\\
[-2mm]\label{a.2}\\[-2mm]
\nu&=&\frac{\alpha}{2}\left(1+\gamma\right)+\frac{\beta}{2}\left(1+\gamma\right)+\frac{\gamma}{2}+\delta+\eta.\nonumber
\end{eqnarray}

By using the Frobenius method \cite{arf}, we can obtain a polynomial solution to the confluent Heun equation. Let us write the confluent Heun function as a power series around the origin, i.e., $H\left(x\right)=\sum_{k=0}^{\infty}a_{k}\,x^{k}$. Thereby, from Eq. (\ref{a.1}) we obtain the relation:
\begin{eqnarray}
a_{1}=-\frac{\mu}{\beta+1}\,a_{0},
\label{a.3}
\end{eqnarray}
and the recurrence relation:
\begin{eqnarray}
a_{k+2}=\frac{\left[\left(k+1\right)\left(k+2+\beta+\gamma-\alpha\right)-\mu\right]}{\left(k+2\right)\left(k+2+\beta\right)}\,a_{k+1}+\frac{\left[\alpha\,k+\mu+\nu\right]}{\left(k+2\right)\left(k+2+\beta\right)}\,a_{k}.
\label{a.4}
\end{eqnarray}

By using Eq. (\ref{a.2}), we can rewrite Eqs. (\ref{a.3}) and (\ref{a.4}) in the form:
\begin{eqnarray}
a_{1}=-\frac{\left[\alpha\left(1+\beta\right)-\beta\left(1+\gamma\right)-\gamma-2\eta\right]}{2\left(\beta+1\right)}\,a_{0},
\label{a.5}
\end{eqnarray}
and
\begin{eqnarray}
a_{k+2}&=&\frac{\left[2\left(k+1\right)\left(k+2+\beta+\gamma-\alpha\right)-\alpha\left(1+\beta\right)+\beta\left(1+\gamma\right)+\gamma+2\eta\right]}{2\left(k+2\right)\left(k+2+\beta\right)}\,a_{k+1}\nonumber\\
[-2mm]\label{a.6}\\[-2mm]
&+&\frac{\left[\alpha\,k+\delta+\frac{\alpha}{2}\left(2+\beta+\gamma\right)\right]}{\left(k+2\right)\left(k+2+\beta\right)}\,a_{k}.\nonumber
\end{eqnarray}

Therefore, from Eq. (\ref{a.4}) or Eq. (\ref{a.6}), the confluent Heun series becomes a polynomial of degree $n$ when we impose two conditions:
\begin{eqnarray}
a_{n+1}&=&0;\nonumber\\
[-2mm]\label{a.7}\\[-2mm]
\delta&=&-\alpha\left[n+\frac{1}{2}\left(2+\beta+\gamma\right)\right],\nonumber
\end{eqnarray}
where $n=1,2,3,4,\ldots$. Unfortunately, we do not know if a closed expression for the asymptotic behaviour of the confluent Heun function for large values of its argument exists. For this reason, we have used the information given from Eq. (\ref{a.3}) to Eq. (\ref{a.7}) in this work with the purposed of analysing each coefficient $a_{k}$ separately.

\acknowledgments{The authors would like to thank the Brazilian agencies CNPq and CAPES for financial support.}

\end{document}